# Towards understanding the behavior of polyelectrolyte – surfactant mixtures at the water / vapor interface closer to technologically-relevant conditions

Sara Llamas,[1] Laura Fernández-Peña,[1] Andrew Akanno,[1,2] Eduardo Guzmán,[1,2,*] Víctor Ortega,[2] Francisco Ortega,[1,2] Aurelio G. Csaky,[2] Richard A. Campbell,[4] and Ramón G. Rubio[1,2,*]

[1] Departamento de Química Física I-Universidad Complutense de Madrid, Ciudad Universitaria s/n, 28040 Madrid, Spain

[2]Instituto Pluridisciplinar-Universidad Complutense de Madrid, Avda. Juan XXIII, 1, 28040 Madrid, Spain

[3] Institut Laue-Langevin, 71 avenue des Martyrs, CS 20156, 38042 Grenoble, Cedex 9, France

*To whom correspondence should be sent: eduardogs@quim.ucm.es and rgrubio@quim.ucm.es

PDADMAC+SLMT at air / water interface

**Abstract**

Polyelectrolyte – surfactant mixtures and their interactions with fluid interfaces are an important research field due to their use in technological applications. Most of the existing knowledge on these systems is based on models in which the polyelectrolyte concentration is around 50 times lower than that used in commercial formulations. The present works marks a step to close the gap on the understanding of their behavior under more practically-relevant conditions. The adsorption of concentrated mixtures of poly(diallyldimethyl)ammonium chloride and sodium N-lauroyl-N-methyltaurate at the water / vapor interface with a crude mixing protocol has been studied by different surface tension techniques, Brewster angle microscopy, neutron reflectometry, and various bulk characterization techniques. Kinetically-trapped aggregates formed during mixing influence the interfacial morphology of mixtures produced in an equilibrium one-phase region, yet fluctuations in the surface tension isotherm result depending on the tensiometric technique applied. At low bulk surfactant concentrations, the free surfactant concentration is very low, and the interfacial composition matches the trend of the bulk complexes, which is a behavior that has not been observed in studies on more dilute mixtures. Nevertheless, a transition to synergistic co-adsorption of complexes and free surfactant is observed at the higher bulk surfactant concentrations studied. This transition appears to be a special feature of these more concentrated mixtures, which deserves attention in future studies of systems with additional components.

## 1. Introduction

The physico-chemical behavior of polymer – surfactant mixtures in the bulk solution and upon adsorption at interfaces, both fluid and solid, has attracted considerable interest in recent years due to their importance in many technological and industrial applications, ranging from drug delivery systems to mineral processing, and from tertiary oil recovery to the development of cosmetic formulations for hair care.[1-6] Among these applications, the mixtures involving polyelectrolytes and surfactants bearing opposite charges are generally accounted for as the most important.[2, 6] In this case, the applications take advantage of the chemical nature of the components, structural features of the formed complexes in solution and the rich phase diagram resulting from the formation of macroscopic aggregates.[1, 3] Despite the extensive research on these systems, many aspects remain not fully understood, especially those in which the interaction of the mixtures with fluid interfaces is considered. This is in part due to the complex interplay of interactions, both electrostatic and hydrophobic, existing in this type of systems, which leads to the appearance of complex phase diagrams and makes their adsorption at fluid interfaces difficult to understand.[4, 7-15] Historically these systems were studied extensively using surface tension measurements by the groups of Goddard[16-17] and Langevin.[18-20] However, the seminal works on the compositional characterization of oppositely charged polyelectrolyte – surfactant layers at the water / vapor interface were performed by Thomas and Penfold who also exploited neutron reflectometry.[21-23] These works represented an important step forward in the understanding of the role of different parameters on the interfacial behavior of polyelectrolyte – surfactant mixtures at chemical equilibrium. More recently, much attention has been paid by Campbell and Varga to linking the interfacial properties of these systems to the bulk phase behavior including non-equilibrium effects both in the bulk and at the water / vapor interface.[24-28] They used the same two techniques in addition to other complementary tools such as ellipsometry and Brewster angle microscopy as well as a range of bulk characterization techniques.

As hinted above, a key characteristic of these systems is their interesting bulk phase behavior.[29-30] At bulk compositions where the formed complexes have a low charge, they spontaneously aggregate over time, and eventually precipitate separates from a depleted supernatant that contains small ions by sinking to the bottom or floating to the top of the samples – these compositions mark the equilibrium two-phase region.[31-32] With a greater amount of either component, the formed complexes have sufficient colloidal stability that they do not spontaneously aggregate, and these compositions mark the equilibrium one-phase regions. An important complication is related to concentration gradients present when the components are mixed, because if complexes formed locally in the sample experience a lack of colloidal stability, aggregates can form on a sub-second time scale that may be insoluble once mixing is complete a few second later.[33] As a consequence, the properties of the mixtures depend critically on the protocol used for mixing the components,[34-35] and this has evidently led to contradictory results in experiments published over the years that introduce further difficulties into predicting their behavior. Moreover, the mixing protocol can be different for different industrial applications, which is an issue that is taken into account in the present work. It has been shown systematically over the last few years that the kinetically-trapped aggregates can affect or even dominate the measured interfacial properties.[25-27] Seemingly in contrast to these experimental studies, however, most of the theoretical studies published so far are based on a thermodynamic framework at chemical equilibrium such as mean field approaches,[36] which do not take into account effects of kinetically-trapped aggregates that can dominate the actual measured steady state properties.[34-35]

Surface tension is probably the most common property used for the study of the adsorption of these mixtures at the water / vapor interface. In most cases polyelectrolytes are not surface active except at high concentrations. However, two key non-equilibrium characteristics of these systems have been systematically described:[24] the slow approach to equilibrium limited by the time scale of precipitation (i.e. aggregate growth and settling), and a perturbation from equilibrium as a result of the sample handling when formed precipitate encounters the water / vapor interface (i.e. when material is spread to form a kinetically-trapped film). In these cases, steady state properties can be measured that are far from equilibrium. Nevertheless, even though the surface tension is formally an equilibrium property, for the sense of simplicity hereinafter we will refer to the steady state property measured in the present work as surface tension.

Mixtures formed by a polycation, poly(diallyldimethylammonium chloride) (PDADMAC) and the anionic surfactant sodium dodecyl sulfate (SDS) at the water / vapor interface are one of the most studied in recent years. This system shows a non-monotonic dependence of the surface tension on the bulk surfactant concentration (at fixed bulk polyelectrolyte concentration), exhibiting a surface tension peak in its isotherm.[14] The maximum of this peak has been shown to be positioned on the phase boundary (i.e. the edge of the equilibrium two-phase region that has the lowest bulk surfactant concentration).[25] The peak cannot be described in terms of simple adsorption models such as Langmuir or Frumkin, which are based on all the components of the system being in the same phase.[37]

PDADMAC has also been chosen as the polycation in this work because it is used as a conditioning agent in hair cosmetic products, flocculant agent for water treatment, retention agent in paper fabrication, and the fabrication of polyelectrolyte multilayers.[38-41] For the anionic surfactant, we have chosen sodium N-lauroyl-N-methyltaurate (SLMT), as it is less susceptible to the degradation into highly surface-active impurities than SDS.[42] In a previous work we have studied the adsorption of this mixture at water / solid interface for the evaluation of its potential applicability as hair conditioning system because it is less irritant than the sodium lauryl sulfonate that is present in most of the current commercial shampoo formulations.[39, 43] The present study extends the scope of the previous one to understanding the behavior of such mixtures at water / vapor interface. The surface tension isotherm has been measured using five different types of tensiometer in order to identify then appropriately categorize any experimental artifacts that could affect the measured values.[44-46] In addition to the study of the surface tension, neutron reflectometry experiments have been carried out to obtain independently a quantification of the interfacial composition of the mixtures, and Brewster angle microscopy has been used to image the presence of kinetically-trapped aggregates at the interface. To the best of our knowledge, this study is the first one in which a combination of different tensiometric techniques, neutron reflectometry and Brewster angle microscopy experiments are used for the study of the interfacial properties of a polyelectrolyte – surfactant system.

Most of the papers published so far have focused on mixtures with polyelectrolyte concentrations well below the overlapping concentration, c*, i.e. the concentration above which the physico-chemical properties of polyelectrolytes lose their dependence on the molecular weight,[47-48] which is far from the situation found in most technological applications. For example, it is estimated that when a shampoo is applied to the wet hair, the polyelectrolyte concentration is about 5 – 10 times c* for PDADMAC, which is around 0.08 wt%.[39, 49-51] As a step towards more industrially-relevant formulations, the bulk polyelectrolyte concentration used in the present work is 0.5 wt%, which is at least an order of magnitude greater than the concentrations used in fundamental studies in the literature.[21, 25] In addition, the samples are all pH-adjusted and a high ionic strength of added inert electrolyte is used. The equilibrium two-phase region would be expected to be at greater bulk surfactant concentrations than those measured, and this is an assumption that is objectively tested. Nevertheless, it is understood that effects from kinetically-trapped aggregates formed during mixing can affect interfacial properties in the equilibrium one-phase regions,[26-27] so this possibility is also examined. The critical mixing of the components is carried out by diluting concentrated samples of pre-mixed components. This approach would be expected in fact to encourage the production of kinetically-trapped aggregates in the samples,[34-35] which is intentional, as rather than to resolve the equilibrium properties, our approach is provide a bridge between the fundamental studies on dilute systems and access the composition and conditions closer to technologically-relevant mixtures.

## 2. Experimental Section

**2.1 Chemicals.** Ultrapure deionized water used for cleaning and solubilization had a resistivity higher than 18 MΩ·cm and a total organic content lower than 6 ppm (Younglin 370 Series, South Korea). PDADMAC was purchased from Sigma-Aldrich (Germany), and had an average molecular weight in the 100 – 200 kDa range, and was used without further purification. SLMT and perdeuterated SLMT were synthetized according to the procedure described in Appendix A; for this purpose either hydrogenous lauric acid or perdeuterated lauric acid ($d^{23}$-lauric acid) was used, purchased from TCI Europe (Belgium) and Cambridge Isotopes (United Kingdom), respectively. Other precursors used were the sodium salt of methyl-taurine (TCI Europe, Belgium) and oxallyl chloride (Acros). Hydrogenous and perdeuterated SLMT were purified by recrystallization from 2-propanol. For the neutron reflectometry experiments, deuterated water ($D_2O$, isotopic purity > 99.9 wt%) was purchased from Sigma-Aldrich (Germany).

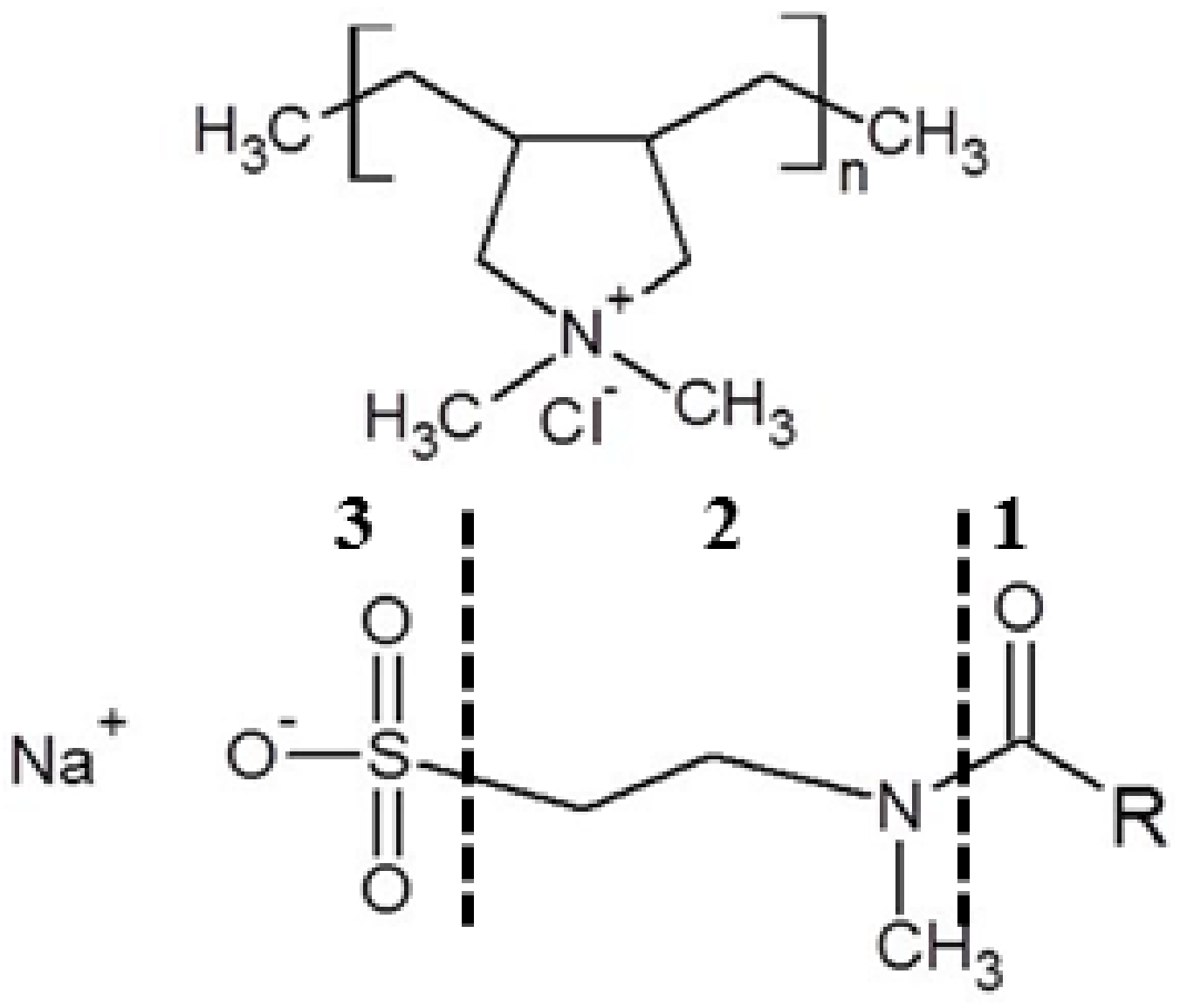

**Scheme I.** Molecular structures of PDADMAC and SLMT with R being an hydrocarbon chain containing 11 carbon atoms. The vertical dashed lines on the surfactant show how the molecule needed to be split up into different layers for the structural analysis using neutron reflectometry.

The pH of all solutions was adjusted to 5.6 using glacial acetic acid (purity > 99 %) and the ionic strength was kept constant by adding 0.3 wt% of KCl (purity > 99.9 %). The PDADMAC concentration was kept constant at 0.5 wt% in all the samples. The SLMT concentration, [SLMT], is defined as the ratio between the mass of surfactant, $m_S$, and the total mass of the solution, $m_T$, and varies from $10^{-9}$ to $10^{-3}$ g/g.

**2.2. Samples preparation.** Polyelectrolyte – surfactant mixtures were prepared in all the cases by weight as it is described in the following: first the required amount of PDADMAC from an aqueous stock solution with concentration 20 wt% is poured into a flask to prepare final solutions with polymer concentration of 0.5 wt%. Afterwards the required amount of the KCl is added until to reach a salt concentration in the final sample of 0.3 wt%. The last step involves the addition of the surfactant solution (pH ~ 5.6) and the final dilution with acetic acid solution of pH ~ 5.6 to reach the desired bulk compositions. In all the cases, the added surfactant solution presents a SLMT concentration, in g/g, one order of magnitude higher than that required in the final sample. The addition of the different component was performed without any delay time between the addition of the different components. Once the samples were prepared, they were mixed by mild stirring (1000 rpm) using a magnetic stirrer during one hour. This procedure inevitably produced kinetically-trapped aggregates during the initial mixing, and their persistent after dilution as well as their effects on the interfacial properties are discussed. In order to ensure reproducibility in the results all the samples were aged during one week before they were used in measurements.

**2.3 Techniques**

*a. Surface tension measurements*

The surface tension of SLMT and PDADMAC – SLMT mixtures was measured against the bulk surfactant concentration using different tensiometric techniques and tensiometers. The adsorption at the water / vapor interface was measured until steady state was reached, i.e. changes of surface tension smaller than 0.1 $mN·m^{-1}$ during 30 minutes. Special care was taken to minimize evaporation effects. For this purpose, the different tensiometers were mounted inside a closed methacrylate box with high relative humidity obtained using open flasks containing saturated solutions of $K_2SO_4$. In addition, whenever possible, special closed cells were used. All experiments were carried out at 25.0 ± 0.1ºC.

*a.1. Surface force tensiometers*. A surface force tensiometer from Krüss K10 (Germany) was first used with different types of Pt contact probes: a Wilhelmy plate and a de Nöuy ring. Special care was taken for the cleaning of the Pt probes, which were burned after each experiment to ensure the removal of any sign of organic materials that could be adsorbed onto its surface as any adsorption would modify plate wettability thus reducing the reliability of the results.

Additionally, a surface force tensiometer from Nima Technology (United Kingdom) was used with disposable paper plates (Whatman CHR1 chromatography paper), ensuring a zero contact angle. A new paper plate was used for each measurement to avoid potential modifications on the plate surface due to the adsorption of material.

Each reported experimental data point was the average of at least three independent measurements. The precision of $\gamma$ was $\pm$ 2 $mN.m^{-1}$. The time of the experiments was long enough to ensure that steady state had been reached.

*a.2. Shape profile analysis tensiometer.* The surface tension of PDADMAC – SLMT mixtures at the water / vapor interface was also measured using a home-made profile analysis.[46, 52] Samples were measured using a pendant drop. In addition, some samples were measured using a bubble configuration. It is important to keep in mind that while the drop hangs in air, the bubble is surrounded by a much larger volume of the solution.

*b. Brewster Angle Microscopy*

Brewster Angle Microscopy (BAM) images were obtained with an $EP^3$-Imaging Ellipsometer from Nanofilm (Germany) at an angle of 53.1° and using a Nd-YAG laser (50 mW) with $\lambda$ = 532 nm. Images collected through a 10x objective (Nikon, Japan) were collected using a CCD camera. Before each measurement the focus was corrected automatically using a software controlled objective. The height of the interface was modified manually to ensure that the images were obtained directly from the interface.

*c. Neutron reflectometry*

PDADMAC+SLMT at air / water interface

Neutron reflectometry experiments were carried out using the time-of-flight horizontal reflectometer FIGARO at the Institut Laue-Langevin (ILL, Grenoble, France).[53] The neutron reflectivity profiles of $\log_{10}R(Q)$ were recorded in the range of wavelengths $\lambda$ = 2 – 30 Å to ensure good overlap between the data measured at incident angles $\theta$ = 0.622° and 3.78°, where R is the measured reflectivity and $Q = \frac{4\pi}{\lambda}\sin\theta$ is the momentum transfer in the range 0.005 – 0.4 $Å^{-1}$. Data were normalized with reference to a measurement of pure $D_2O$. The analysis was carried out using Motorfit,[54] which is based on the application of an optical matrix model based on Abele's formulism[55] to the experimental data. Mixtures of PDADMAC with deuterated surfactant were measured in pure $D_2O$ and in air contrast matched water (ACMW), which is a mixture of 8.1% v/v $D_2O$ in $H_2O$ to give zero scattering length density. Two different data analysis approaches were used.

First, a direct measure of the SLMT surface excess ($\Gamma_S$) in 10 mixtures was performed by fitting the scattering excess at low-Q values (0.01 – 0.03 $Å^{-1}$) of data recorded in only the isotopic contrast involving ACMW. Background was not subtracted from the data. In this case, the contribution from the polyelectrolyte was neglected as the scattering length of one of its monomer unit ($b_P$ = 2.7 fm) is negligible compared with that of one deuterated surfactant molecule ($b_S$ = 262 fm). In this model, the layer thickness (d) was fitted with an arbitrary scattering length density of $\sigma = 3 \times 10^{-6}$ $Å^{-2}$, fixed roughness values of 4 Å, and a background of $5.65 \times 10^{-5}$ fitted from measurements of pure ACMW. The following equation was then solved:

$$\Gamma = \frac{\sigma d}{b_S N_A} \quad (1)$$

where $N_A$ is Avogadro's number.

PDADMAC+SLMT at air / water interface

Second, a structural analysis was carried out by applying a consistent physical model to data recorded in both isotopic contrasts over the whole accessible Q-range. The two isotopic contrasts were chosen because the one in ACMW is most sensitive to the amount of surfactant and the one in $D_2O$ is most sensitive to the amount of polyelectrolyte. Background was subtracted from the data using the area detector. In a first step, data of pure deuterated SLMT at its critical micelle concentration were fitted using a three-layer model corresponding to the divisions on the molecule marked in scheme 1. While it is not common to have to split the surfactant into 3 layers, the necessity for this arises because of the sharp variations in scattering length density along the length of the molecule: the first portion with deuterated chains has a high scattering length density ($6.8 \times 10^{-6}$ Å$^{-2}$), the second portion with the organic part of the head group has a low scattering length density ($0.28 \times 10^{-6}$ Å$^{-2}$), and the third portion with the ionic part of the head group has an intermediate scattering length density ($3.0 \times 10^{-6}$ Å$^{-2}$). As such, a two-layer model was inadequate to fit the data. The thickness of the chains layer was fitted with a volume fraction of 1, and in order to conserve the same number of the parts of the surfactant molecules the volume fractions of the two head group layers ($\varphi_{2,3}$), each fixed at a thickness of 3 Å, were constrained according to

$$\varphi_{2,3} = \frac{\sigma_1 d_1 b_{2,3}}{\sigma_{2,3} d_{2,3} b_1} \tag{2}$$

Fixed roughness values of 4 Å and a residual background of $3 \times 10^{-7}$ were used. In a second step, data of PDADMAC – SLMT mixtures were fitted where polyelectrolyte was incorporated into the lower head group layer in contact with the solvent. Other structural models were attempted but this one gave the best fits to the data. In this case, the thicknesses of the chains and lower head group layers were fitted, the volume fraction of the upper head group layer was constrained according to eq. 2, and the volume fraction of the lower head group layer was constrained again so that there was a consistent number of each part of the surfactant molecule according to

$$\varphi_3 = \frac{\sigma_1 d_1 b_3 \left(\sigma_{heads,3} - \sigma_{poly}\right)}{\sigma_3 d_3 b_1 \left(\sigma_{fitted,3} - \sigma_{poly}\right)} \quad (3)$$

The fitted surfactant surface excesses resulting from the two data analysis approaches were in good agreement as is demonstrated below.

*d. Electrophoretic mobility measurements*

Electrophoretic mobility, $\mu_e$, measurements were carried out using a Zetasizer Nano ZS instrument (Malvern Instruments Ltd., Worcestshire, UK).

*e. Turbidimetry*

The turbidity of the samples was determined by measuring the transmittance of the mixtures at 400 nm using a UV/visible spectrophotometer (HP-UV 8452). The turbidity is represented as (100-T (%))/100, with T being the transmittance.

## 3. Results and discussion

### 3.1. Adsorption of the pure components at the water / vapor interface

Understanding the adsorption of PDADMAC – SLMT mixtures at the water / vapor interface requires a previous understanding of the adsorption of the two components separately. Figure 1 shows the surface tension isotherms obtained for the adsorption of PDADMAC and SLMT separately at different bulk concentrations (note that all solutions were made in 0.3 wt% of KCl at pH 5.6).

PDADMAC+SLMT at air / water interface

There is a negligible adsorption of PDADMAC at the water / vapor interface, as γ is equivalent to the value of pure water over the whole concentration range studied. These results are in good agreement with the results by Noskov et al.[56] who showed that PDADMAC is not surface active for concentrations lower than 3 wt%. For SLMT, γ decreases as the bulk surfactant concentration increases until the concentration overcomes the critical micellar concentration (cmc). The surface tension values for the deuterated SLMT coincide with those of its hydrogenous analogue (data not shown).

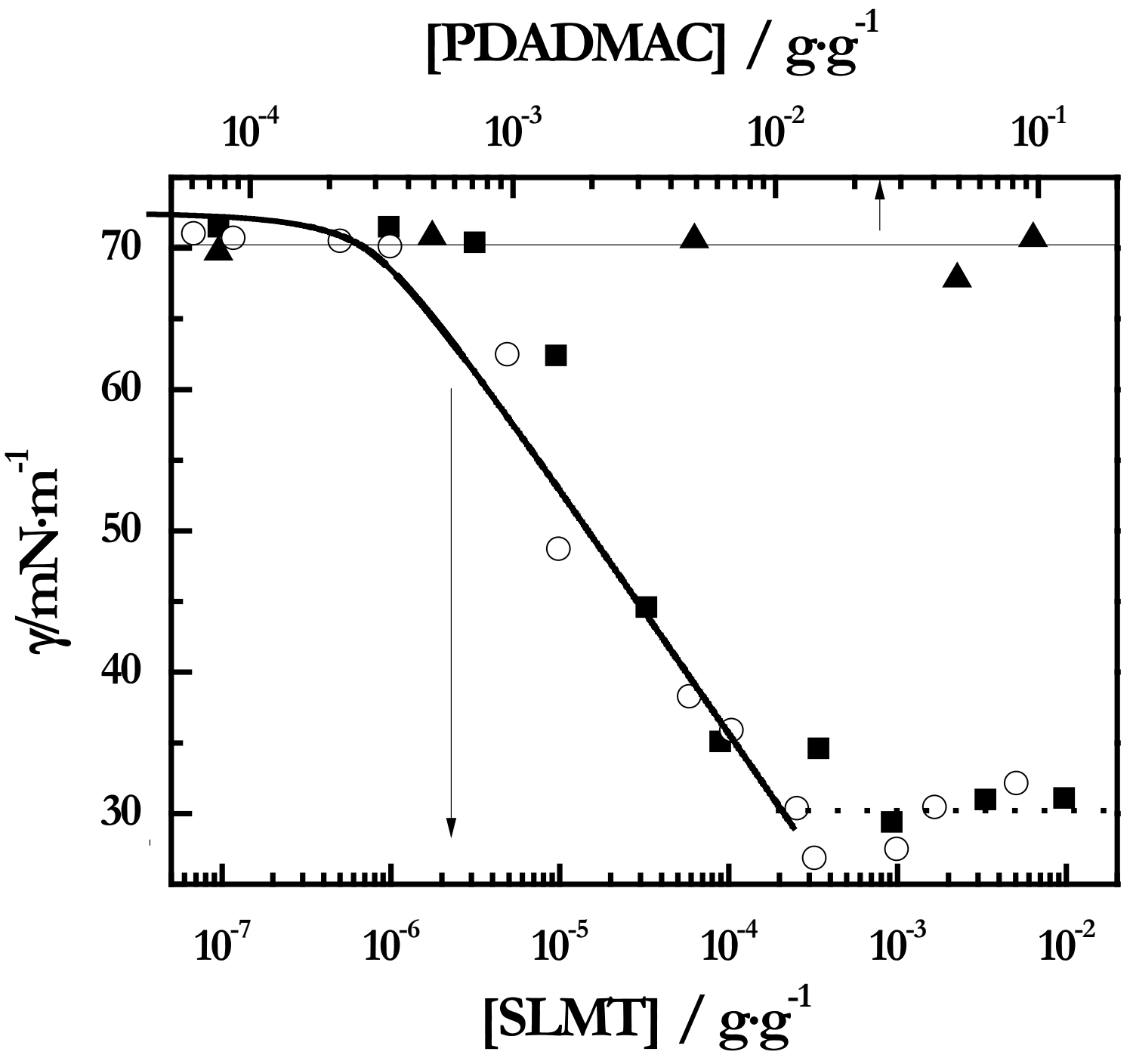

**Figure 1.** Surface tension dependence on PDADMAC concentration for pure PDADMAC solutions (top axis, ▲). The line is a guide for the eyes. Surface tension dependence on SLMT concentration (botton axis) for pure SLMT solutions as was obtained using a Wilhelmy plate tensiometer with a Pt probe (○) and a drop shape analysis tensiometer (■). The solid line represents the best fit of the experimental isotherms to Frumkin isotherms, whereas the dotted line represents the region of quasi-constant surface tension once the cmc is overcome.

PDADMAC+SLMT at air / water interface

The γ data obtained with the different tensiometers agree with each other within the combined errors, although the results of the drop shape tensiometer are slightly higher than those of the Pt Wilhelmy plate tensiometer. This can be understood in terms of the partial depletion of surfactant from the bulk due to the adsorption at the interface. The depletion is important in drop profile analysis tensiometer experiments because the ratio of the interfacial area to the drop volume is much higher than for the Pt Wilhelmy plate tensiometer. This surfactant depletion in the drop has also been discussed very recently by Fainerman et al.[57] Therefore, a slight depletion of the bulk concentration associated with the surfactant adsorption at the interface will present a more critical effect in the case of the drop measurements. It is worth mentioning that surface tension measurements provide a value of the cmc about $2 \times 10^{-4}$ g/g for SLMT. The experimental isotherm for SLMT (see Figure 1) can be described using the Frumkin isotherm defined as[37]

$$bc = \frac{\Gamma\omega}{1-\Gamma\omega} e^{-2a\Gamma\omega} \tag{4}$$

$$-\frac{\Pi\omega}{RT} = \ln(1-\Gamma\omega) + a\left(\Gamma\omega\right)^2 \tag{5}$$

where c is the bulk concentration whereas Γ is referred to the surface excess, ω is the area of a molecule in the close packed monolayer, b corresponds to the constant of the adsorption equilibrium, *a* is the parameter of molecular interaction, and R and T are the gas constant and the absolute temperature, respectively. Table 1 summarizes the parameters for the best fit to the Frumkin model of the combined data obtained by both tensiometers used.

**Table 1.** Parameters obtained from the best fit of the experimental isotherm for SLMT to a Frumkin model (Figure 1)

| | $10^{-2}$·b (l/mmol) | $10^{-5}$·ω (m$^2$/mol) | *a* |
|---|---|---|---|
| PDADMAC+SLMT at air / water interface | | | |

| | | |
|---|---|---|
| 1.36 | 3.26 | 1.32 |

**3.2. Interactions of PDADMAC – SLMT mixtures in the bulk solution**

A necessary first step before a study of the adsorption of PDADMAC – SLMT mixtures at fluid interfaces is an analysis of the charge of the formed complexes and an examination of any aggregation that takes place in the bulk. Figure 2 shows the dependence of the electrophoretic mobility and turbidity on the bulk surfactant concentration.

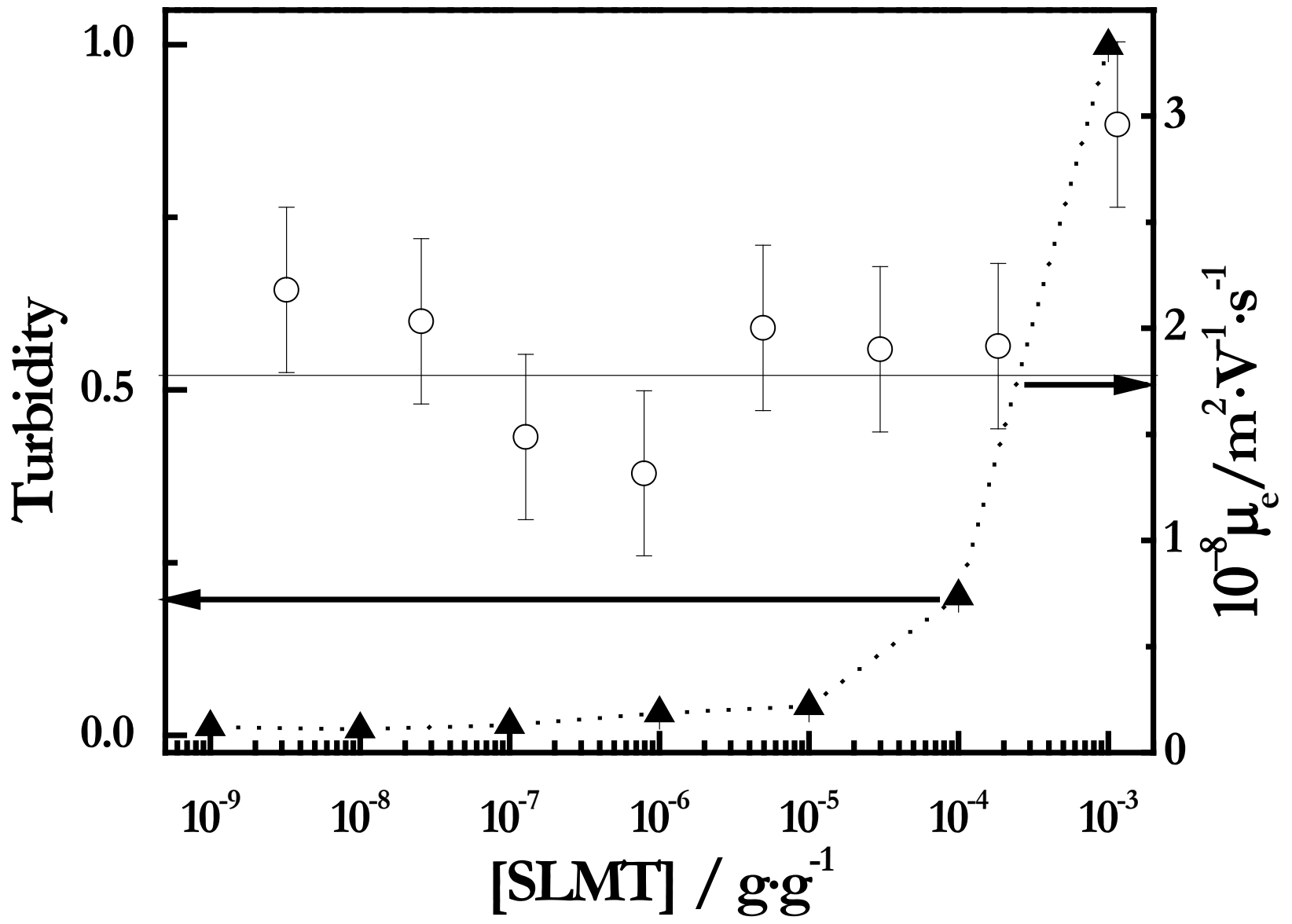

**Figure 2.** SLMT concentration dependences of the turbidity (left axis, ▲) and electrophoretic mobility (right axis, ○) for PDADMAC – SLMT mixtures. The solid line is referred to the electrophoretic mobility value of PDADMAC. The results correspond to samples aged during one week.

The results of the electrophoretic mobility show positive values, close to that corresponding to the polyelectrolyte, over the entire range of bulk surfactant concentrations studied. Thus, it is possible to infer that the samples all contain complexes with excess polyelectrolyte segments. This finding can be rationalized with reference to the efficient binding and low free surfactant concentration at charge neutralization (~ 0.2 mM) for 100 ppm PDADMAC – SDS mixtures. [17,18] If one neglects any differences in the binding isotherm from the change of surfactant structure and higher ionic strength, and if little variation in the free surfactant concentration at charge neutralization is assumed with changing bulk polyelectrolyte concentration, a simple extrapolation of the bulk polyelectrolyte concentration by a factor of 50 would put charge neutralization at a higher bulk surfactant concentration than the samples measured in Figure 2. Hence it follows that the complexes in all the samples measured are in the undercharged regime.

This scenario indicates that all the mixtures studied fall in an equilibrium one-phase region where freshly mixed samples are expected to be optically transparent. Nevertheless, at compositions close to the phase boundary, kinetically-trapped particles can be produced during mixing, and this results in turbid samples.[33] Therefore it is helpful to examine the phase behavior of the samples by carrying out turbidity measurements. Only for the higher bulk SLMT concentrations studied, there is the onset of phase separation as evidenced by the increase of the turbidity. This is an essential difference in relation to many studies on more dilute systems in the literature, in which the surface tension dependence on the bulk surfactant concentration covered a transition from undercharged through neutral to overcharged complexes.[21-22,24,25, 58] As there is no indication of the approach of charge neutralization of the bulk complexes from the electrophoretic measurements measurements in the samples in Figure 2, we may infer that the turbidity arises from the presence of kinetically-trapped aggregates that were produced as the components were mixed then diluted to the stated bulk composition.[34-35]

### 3.3. Adsorption of PDADMAC – SLMT mixtures at the water / vapor interface

PDADMAC+SLMT at air / water interface

Measurements performed using different tensiometers can help to shed light on the complex interfacial behavior present in oppositely charged polyelectrolyte – surfactant solutions. Figure 3a shows the surface tension obtained by a drop shape tensiometer. For the sake of comparison the data corresponding to the SLMT is also included.

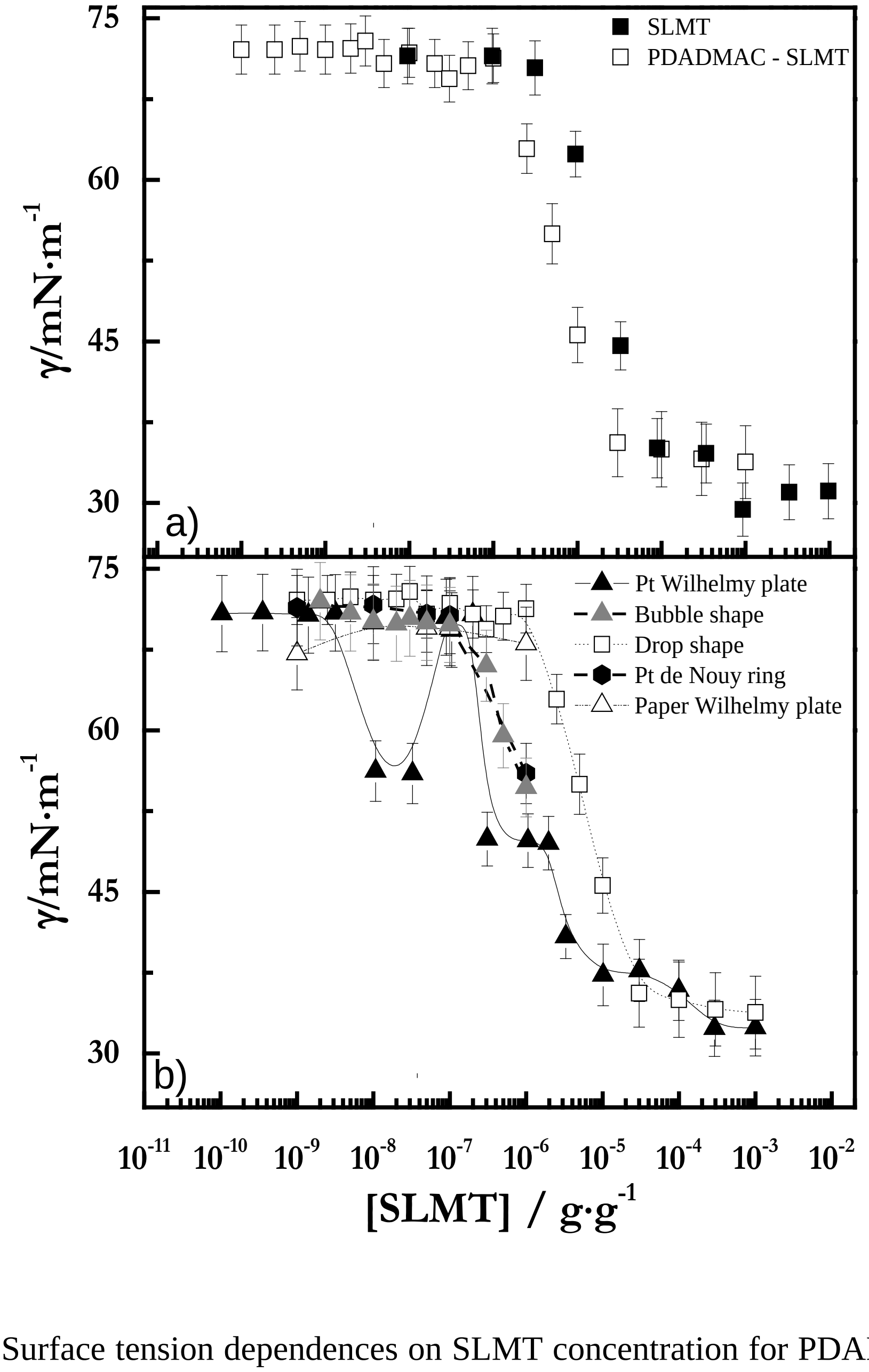

**Figure 3.** (a) Surface tension dependences on SLMT concentration for PDADMAC – SLMT mixed solutions and pure SLMT solutions measured with a drop shape tensiometer. (b) Surface tension dependences on the bulk SLMT concentration for PDADMAC – SLMT PDADMAC+SLMT at air / water interface

mixtures obtained using five different tensiometers. The errors bars were calculated as the standard deviation of the results obtained from three independent measurements. The results correspond to samples aged during one week.

For the lowest bulk surfactant concentrations measured, the behavior is similar for both pure SLMT and PDADMAC – SLMT mixtures, so it could be assumed that only a low coverage of molecules adsorb at the fluid interface. The surface concentration is not enough to lower the free energy of the interface, thus the surface tension remains close to the value corresponding to pure water. An increase of the bulk SLMT concentration leads to a monotonic lowering of $\gamma$ for the mixtures, with the decrease equivalent for bulk surfactant concentrations about one order of magnitude lower than for the pure surfactant. This type of behavior has been previously reported for the adsorption of other polymer – surfactant mixtures at the water / vapor interface.[4, 21-22, 24-25, 59]

The data and physical description above do not include any non-regular trends in the surface tension isotherms. In contrast, Figure 3b shows that different results were obtained when different tensiometers were used. Fluctuations in the data are evident only when Wilhelmy plate tensiometer with Pt probe was used. Such a type of discrepancy between the results obtained using different techniques has been previously reported by Noskov et al.[60] This behavior was explained in terms of the following considerations: a) the aggregation kinetics of polyelectrolyte – surfactant complexes in the bulk, and b) the differences on the physical bases of the different techniques.

As already mentioned, oppositely charged polyelectrolyte – surfactant mixtures are systems that are often dominated by non-equilibrium effects related to the production of kinetically-trapped aggregates during mixing of the components. [34-35] The resulting direct effects on the interfacial properties have also been characterized, as it has been shown that fluctuations in the amount of material at the interface of a polyelectrolyte – surfactant mixture were evident in samples more than three orders of magnitude in bulk composition from the equilibrium two-phase region,[27] and that such kinetically-trapped aggregates could be visualized using BAM.[26] Thus, to examine if the appearance of the surface tension fluctuations in PDADMAC – SLMT mixtures for the measurements carried out using the Wilhelmy plate tensiometer with a Pt probe may be associated with the formation of aggregates that can be trapped at the interface and consequently adsorbed onto the Pt probe, we studied the system using BAM (see Figure 4a and 4b). It can be seen that there are aggregates that have coalesced laterally to form isolated islands. As such, given the inhomogeneity in the lateral morphology, to a first approximation one may assume that their effect on the surface tension might be minimal. Recent work on related systems has shown how polyelectrolyte – surfactant aggregates can dissociate and spread material by Marangoni flow across the liquid / vapor interface in the form of a kinetically-trapped film that can lower the surface tension, but this work was conducted on a pure water subphase without a substantial excess of one component in the bulk.[24] Even so, the aggregates trapped at the interface at bulk compositions far from the equilibrium two-phase region, as imaged using BAM, do coincide with the non-monotonic surface tension behavior from the Wilhelmy plate tensiometer with a Pt probe, hence the issue merits further consideration.

The question is then raised why surface tension fluctuations were observed only using this particular tensiometric technique. We may reason that the trapped, hydrophobic PDADMAC – SLMT aggregates can stick more readily onto the rough planar Pt probe surface, thus changing the contact angle of the Pt plate with the interface, and consequently the value of the surface tension determined using this technique is affected (this adsorption can be the origin of the changes on the contact angle of water drops found for the clean plate and the PDADMAC+SLMT at air / water interface plate after the measurements). If this physical picture were correct, to a first approximation an analogous effect using the Pt De Nouy ring could be expected. However, the smaller surface area in contact with the solutions and the fact that it is not rough make the sticking of aggregates less favorable, and consequently the modification of the contact angle, thus providing surface tension values that are monotonic. The absence of surface tension fluctuations was also found when other tensiometric techniques were used. This observation confirms that the trapped aggregates do not spontaneously dissociate to spread a kinetically-trapped film at the interface of solutions with such a large excess of polyelectrolyte (and counterions) in the bulk. When a paper probe was used the solution wets the plate completely and any possible effect of aggregates sticking is much smaller, and in the case of drop or bubble shape tensiometers there is no equivalent probe so there is no such effect. The irregular large aggregates at the water / vapor interface are mostly observed at the lower bulk SLMT concentrations examined (Figure 4a and 4b). At higher bulk SLMT concentrations, the aggregates are smaller and more homogeneous in appearance (Figure 4c and 4d). This observation is consistent with the results by Monteux et al.[61] on the poly(sodium styrene sulfonate) – dodecyltrimethylammonium bromide system.

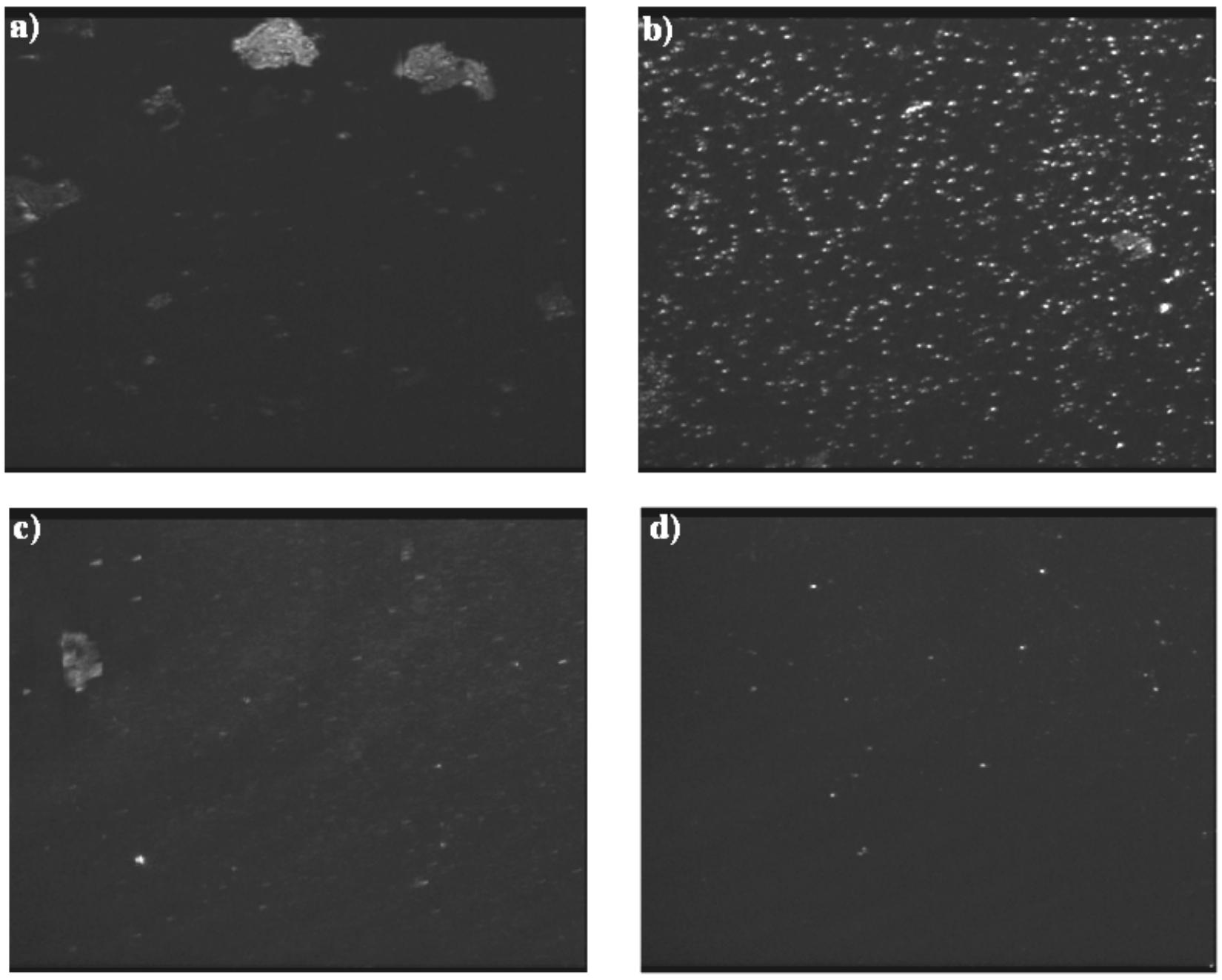

**Figure 4.** BAM images of the lateral morphology of PDADMAC – SLMT mixtures at the water / vapor interface at different values of the bulk SLMT concentration: a) $10^{-9}$. b) $10^{-8}$. c) $10^{-7}$. d) $10^{-6}$ g/g. The dimensions of the images are 438 × 320 $\mu m^2$. The results crrespond to samples aged during one week.

There is one additional difference between the data on PDADMAC – SLMT mixtures recorded using the different tensiometric techniques, one that is systematic rather than random in nature: the data measured using the pendant drop is shifted to higher bulk SLMT concentrations. Discrepancies between the results obtained using drop and bubble profile analysis tensiometers have been found in previous studies.[46, 62-63] Furthermore, previous studies on the adsorption of proteins have evidenced that the results obtained using bubble profile tensiometers agree well with those results obtained by Wilhelmy plate tensiometer.[44] The differences in the isotherms associated with different configurations of analysis profile tensiometer (drop and bubble) can be explained taking into account the different relation between the surface area of the interface and the volume of the solution containing the polyelectrolyte – surfactant mixture. In the case of pendant drop configuration, the adsorption at the surface of the drop leads to a decrease of the effective concentration of the sample solution, i.e., a depletion effect. Thus, the lower volume of the drop in comparison with its surface area leads to a significant decrease of the effective concentration of the bulk, leading to higher values of the surface tension. On the other hand, if the bubble configuration is used, the depletion effect is not observed since the adsorbed amount is negligible in relation to the total concentration of surface-active material in the sample. The same explanation based on the ratio interfacial area / volume of the sample can be used for explaining the discrepancies found between the experiments performed using the drop shape tensiometer, and those obtained using the contact plate/ring probes. The differences associated with the depletion of material in profile analysis tensiometers are schematized in Figure 5.

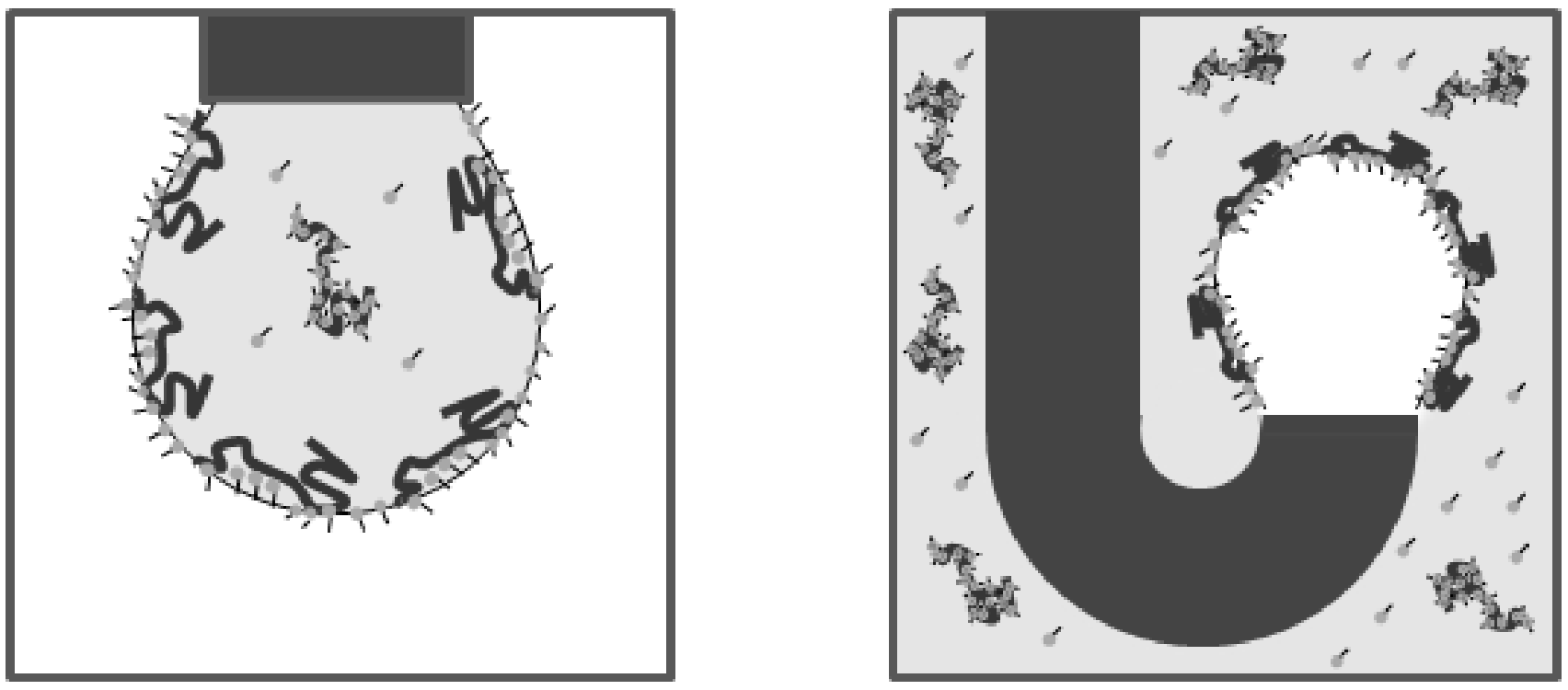

**Figure 5.** Schematic representation of the adsorption of surfactant molecules and polymer – surfactant complexes to the water / vapor interface and the different phenomena that take place depending on the technique used: profile analysis tensiometer in drop configuration (left panel) and in bubble configuration (right panel).

The analysis of the differences between the results obtained for tensiometric techniques involving a different surface area / volume ratio can provide an estimation of the surface excess according to the approximation proposed by Fainerman et al.[57] However, such an approximation is developed for single-component systems, whereas in our case we are dealing with polyelectrolyte – surfactant mixtures. Thus, it is necessary to analyze the degree of binding between the surfactant molecules and the polymer monomers before a calculation of the surface concentration on the basis of the model proposed by Fainerman et al.[57] Figure 6 shows the dependence of the binding degree $\beta = c_s^{free} / c_{monomer}$, with $c_s^{free}$ being the concentration of free surfactant and $c_{monomer}$ the concentration of charged monomers units of the polyelectrolyte, as obtained by potentiometric measurements carried out using surfactant selective electrode.

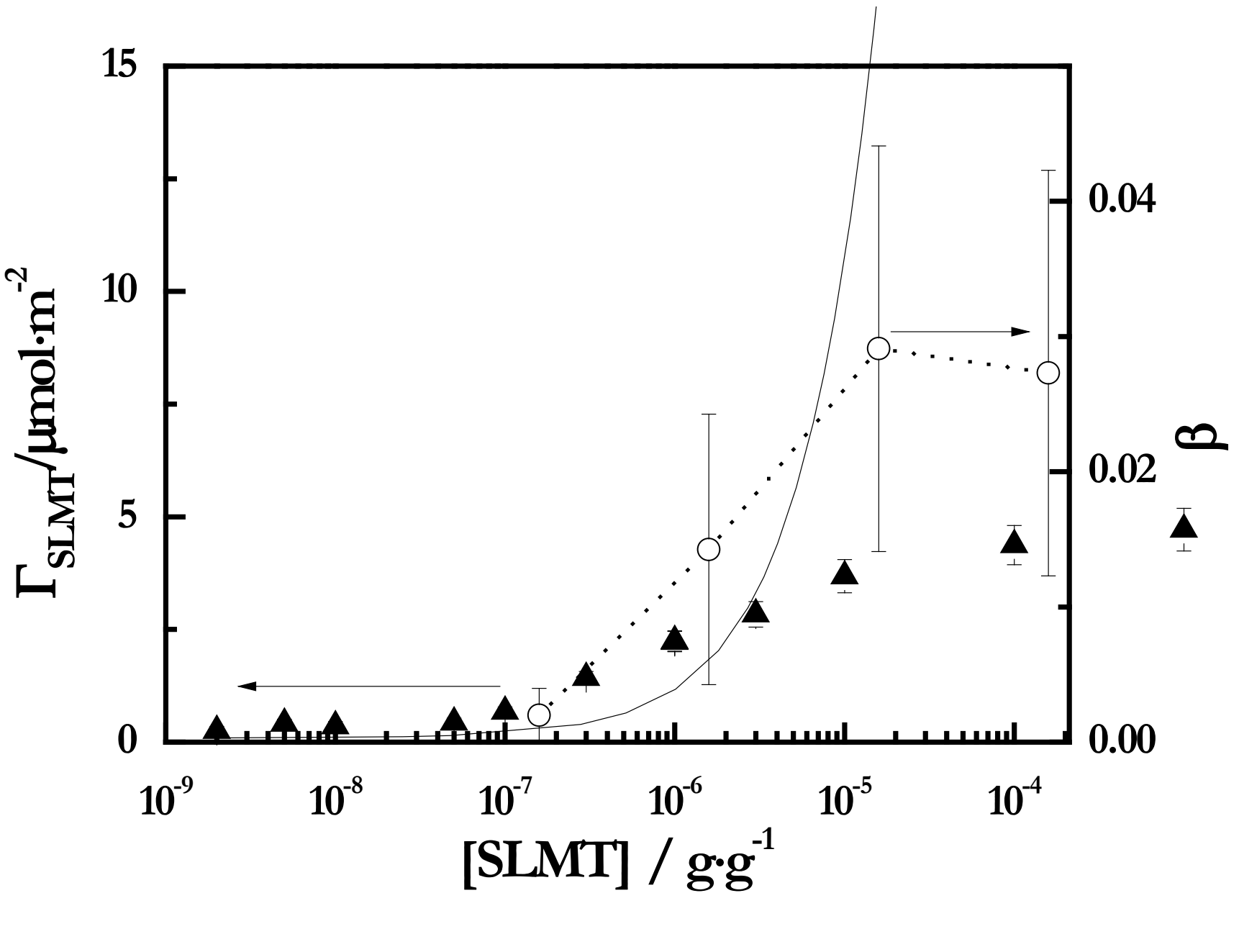

**Figure 6.** Surface excess obtained from eq. 5 (solid line) and surface excess of SLMT (symbols) as was obtained by neutron reflectometry. Note that the error bars are smaller than the data size for the data from neutron reflectometry (left axis, black triangles). Binding isotherm of SLMT on PDADMAC as function of the initial SLMT bulk concentration (right PDADMAC+SLMT at air / water interface

axis, white circles). The line is a guide for the eyes. The results correspond to samples aged during one week.

The binding isotherm evidences a high degree of binding of the surfactant molecules to the polyelectrolyte molecules in the bulk over a broad range of the bulk composition measured. Again, with reference to previous work conducted on much more dilute PDADMAC – SDS mixtures in 0.1 M NaCl,[24-25] and again for simplicity neglecting the difference in the structure of the surfactant and ionic strength, for 100 ppm the binding degree at charge neutralization is ~ 0.3, so an extrapolation of the bulk polyelectrolyte concentration by a factor of 50 without major changes in $c_s^{free}$ places the binding degree at charge neutralization as < 1 %. As all the samples measured in the present study have a much lower bulk surfactant concentrations than charge neutralization and therefore contain undercharged complexes, this would put the binding degree at << 1 %. Indeed the results reveal a very low binding degree, as only around 3 % of the surfactant molecules in the PDADMAC – SLMT mixtures remain free even for the highest concentrations. As a result of this very low binding degree, if we were to make an approximation of zero free surfactant, we could use the surface tension isotherm to estimate the total surfactant surface excess as[57]

$$\Gamma = \frac{V_d}{A_d}\left(c_d - c_b\right)_{\gamma} \qquad (6)$$

where $\left(c_d - c_b\right)_{\gamma}$ is the difference in bulk surfactant concentration for a fixed value of surface tension between the concentration on the drop and the bubble experiments, $V_d$ is the volume of the drop and $A_d$ the area of the drop. Figure 6 shows the surface excess calculated according eq. 6 and the surface excess of surfactant measured directly using PDADMAC+SLMT at air / water interface neutron reflectometry (see first analysis method in the Experimental Section).

The experimental results and the one obtained on the bases of eq. 6 shows a monotonic increase of the surface excess with the surfactant concentration. However, the experimental and calculated values coincide only for the lowest bulk SLMT concentration, where indeed the binding degree is expected to be lowest. This suggests that the complexity of the system at higher bulk surfactant concentrations, where the free surfactant concentration cannot be neglected, requires the development of complex thermodynamic multi-component models that account for the differences on the activities of the different component involved.

In order to deepen our understanding of the composition of the interface, additional neutron reflectometry experiments were carried out. Figure 7 shows an example of data recorded over the whole accessible Q-range in two isotopic contrasts as well as fits using the three-layer model for samples with a bulk SLMT concentration of $3 \times 10^{-6}$ g/g (see second analysis method in the Experimental Section). The inset shows the scattering length density profiles in the two isotopic contrasts with respect to the distance from the interface. Here, effectively the area of the peak in the blue curve is related to the amount of surfactant at the interface, and the area of the inflexion of the green curve is related the corresponding amount of polyelectrolyte. Figure 8 shows the resulting surface excesses of PDADMAC and SLMT obtained from this structural analysis in addition to the 10 SLMT surface excesses in the mixtures repeated from Figure 6.

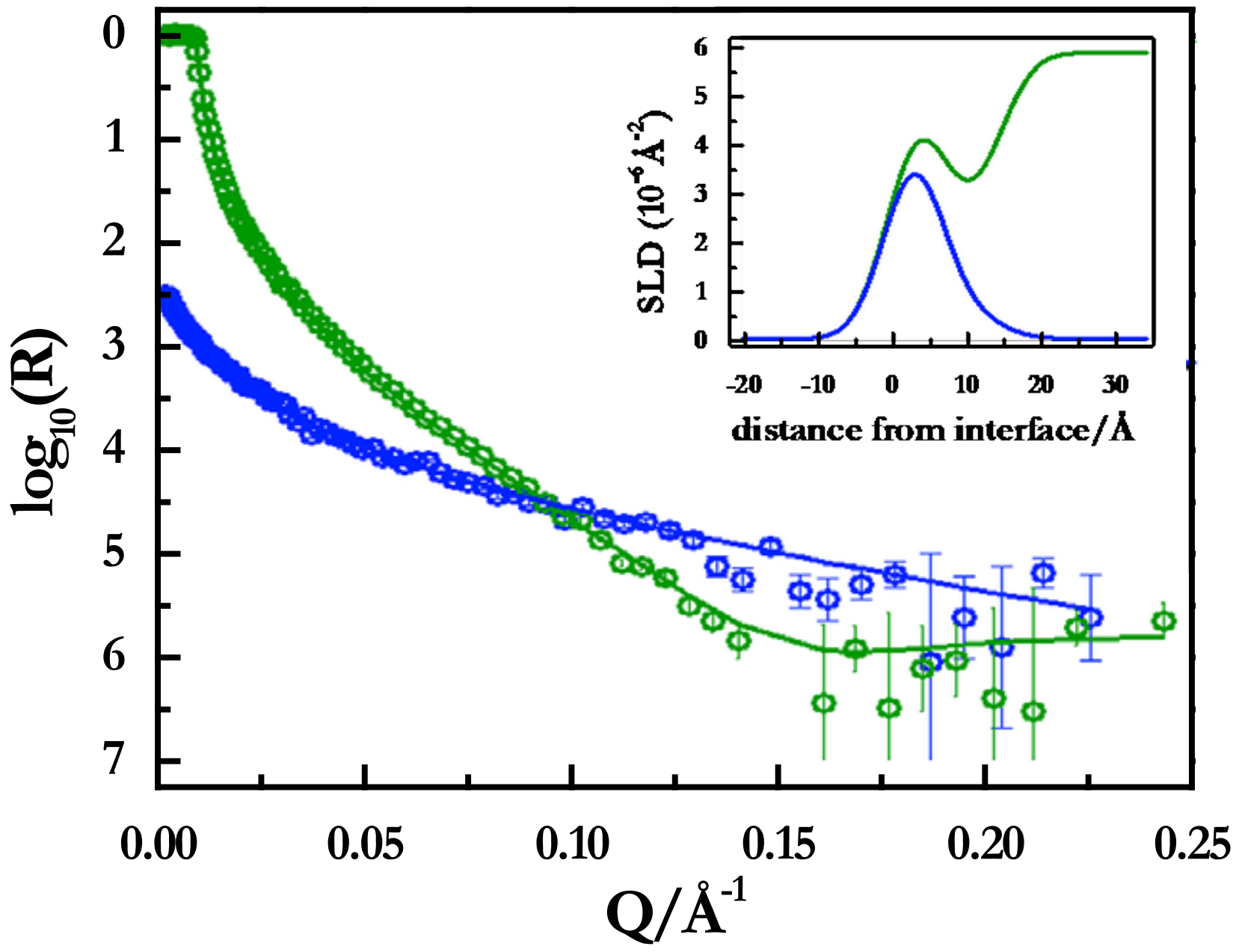

**Figure 7.** Neutron reflectivity profiles of PDADMAC – SLMT mixtures with a bulk SLMT concentration of 3 × $10^{-6}$ g/g. The isotopic contrasts involve deuterated surfactant in ACMW (blue) and in $D_2O$ (green). The model fits are shown as lines. The inset shows the scattering length density (SLD) profiles in the two isotopic contrasts with respect to the distance from the interface. The results correspond to samples aged during one week.

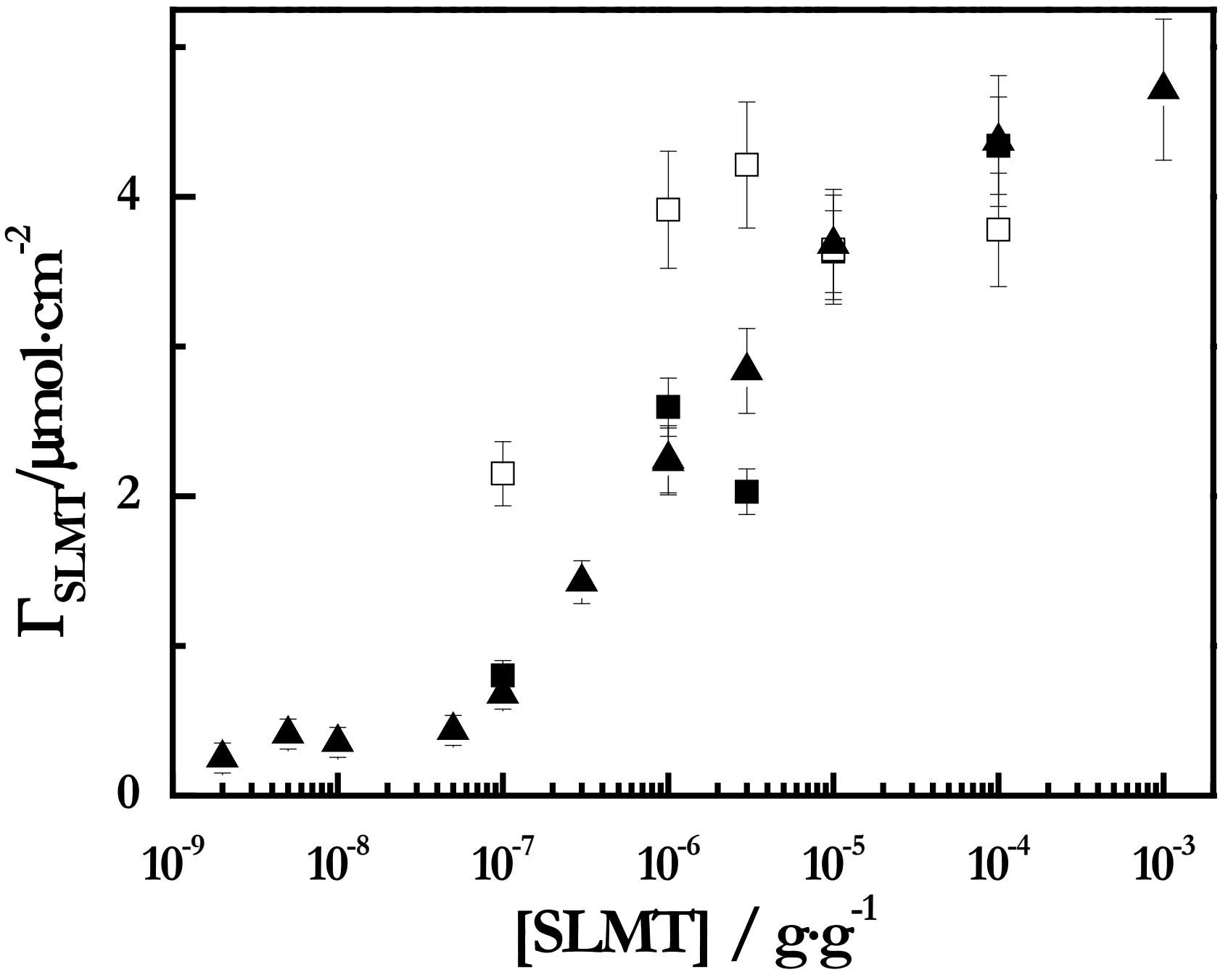

**Figure 8.** Surface excesses of PDADMAC (□) and SLMT (■) in 5 of their mixtures from the structural analysis, and surface excess of SLMT (▲, same results reported in Figure 6) in 10 of their mixtures from the low-Q analysis, both determined using neutron reflectometry. The results correspond to samples aged during one week.

The results show that both methodologies to obtain the surface excess of the surfactant are in good agreement. The use of the structural analysis to determine the amount of PDADMAC at the interface reveals an interesting trend. At low bulk SLMT concentrations, there is a large excess of polyelectrolyte segments at the interface as the polyelectrolyte surface excess (in monomer units) is three times that of the surfactant. However, as the bulk SLMT concentration increases, the interfacial stoichiometry reaches unity by $10^{-5}$ g/g, and at the highest bulk SLMT concentration measure the surfactant surface excess exceeds that of the polyelectrolyte by ~ 15%.

A first interpretation of this trend in the interfacial composition is that qualitatively it mirrors the changes in bulk composition, and that an adequate physical framework to describe the interfacial properties of these samples in the semi-dilute regime, where the free bulk surfactant concentration is relatively low, may be one of adsorption of complexes formed in the bulk. This picture is in contrast to that in studies of samples with undercharged complexes in the dilute regime,[16-18,20] where there is not a significant shift in the interfacial composition with changing bulk composition, which can be attributed to the synergistic co-adsorption of free surfactant with the complexes. The large excess of polyelectrolyte at the interface in the sample with $10^{-7}$ g/g SLMT can be rationalized by the very low free surfactant concentration as a result of the high polyelectrolyte concentration used (see Figure 6). Even so, the complexes are positive charged in all of the samples measured (see Figure 1), and therefore to achieve equimolar binding with a bulk SLMT concentration of $10^{-5}$ g/g, there must be co-adsorption with additional surfactant adsorbed at the interface. This strong effect is evident even though ~ 97 % of the surfactant is bound in complexes with the polyelectrolyte, which underlines the important role of a rather small amount of the free surfactant in determining the interfacial properties of these mixtures. The observed transition between the regimes of complex adsorption and synergistic co-adsorption is clearly an interesting facet of the behavior of the more concentrated samples studied in the present work. This issue merits further investigation in a range of technologically-realistic systems in the future, both in terms of additional components and the behaviour of the systems under dynamic conditions involving flow.

PDADMAC+SLMT at air / water interface

**4. Conclusions**

Our study on the adsorption of mixtures of polyelectrolytes and surfactants bearing opposite charges at the water / vapor interface, involving a bulk polyelectrolyte concentration close to that used in commercial formulations, has evidenced important differences in relation to the numerous studies of more dilute systems. The main origin of this difference lies in the fact that over the range of bulk surfactant concentrations measured, there is a very low free surfactant concentration. This means that a physical framework involving the adsorption of bulk complexes becomes more relevant than in systems with lower bulk polyelectrolyte concentrations in which a framework of synergistic co-adsorption of complexes and free surfactant is more appropriate. Nevertheless, there is stoichiometric binding at the interface even at a bulk compositions far from stoichiometric binding in the bulk, which demonstrates that this synergistic effect can dominate the interfacial behavior even in these more concentrated mixtures when the free surfactant concentration is on the order of just a few percent.

At the same time, even though the samples in the present work existed in an equilibrium one-phase region, and their optical density when freshly mixed remained low except for those with the highest bulk surfactant concentrations measured, the effects of kinetically-trapped aggregates on the interfacial properties were extensive. Our applied mixing protocol of the components (i.e. mix concentrated components then dilute) was deliberately crude in order to mimick the type of stepwise processes used in industrial plants. The resulting particles were imaged at the vapor / air interface at bulk compositions that were several orders of magnitude from the equilibrium two-phase region, and the relatively large size of the islands observed pointed to lateral coalescence of the aggregates at the interface. Further, depending on the tensiometric technique applied, fluctuations were observed in the surface tension isotherm, which was attributed to attachment of the aggregates to the probe. This evidence may help to rationalize some of the numerous discrepanies in the literature from different tensiometric techniques applied to mixed systems.

The main novelty of this work is related to the fact that the system studied is closer not only in composition but also sample history to those used in real technological applications. Thus this information on the interfacial behavior can help to improve our understanding of the physico-chemical bases underlying the performance of polyelectrolyte – surfactant systems in applications involving detergency or foam stabilization. On the basis of this work, it will be important to continue this direction towards mixtures involving a greater a number of components and under dynamic conditions of flow in order to make further steps to a coherent understanding of more relevant mixtures under more realistic conditions.

## Acknowledgements

This work was funded in part by MINECO under grants FIS 2014-62005-EXP and CTQ-2016- 78895-R, by EU under Marie Curie ITN CoWet (Grant Number 607861). This work is based upon work from COST Action MP-1305. The authors are grateful to the CAI of Espectroscopia y Correlación of UCM for the use of their facilities and to the ILL for an allocation of neutron beam time on FIGARO as well as the Partnership for Soft Condensed Matter for access to ancillary equipment during the experiment.

**Appendix A**

**A.1. Synthesis of lauroyl chloride.**

To a solution of lauric acid (5 g, 25 mmol) in anhydrous $CH_2Cl_2$ (50 mL) under Ar DMF (50 μl) was added followed by a 2 M solution of oxalyl chloride in $CH_2Cl_2$ (37.4 mL, 75 mmol). The mixture was stirred at room temperature for 1 h, and then at 40ºC for 3 h. Volatiles were removed under reduced pressure, leaving an oily residue (crude lauroyl chloride, pale yellow oil) which was used in the next step without further purification.

**A.2. Synthesis of sodium N-lauroyl-N-methyltaurate.**

To a stirred solution of sodium *N*-methyltaurate (4.0 g, 25 mmol) in water (20 mL) at 10-15ºC under Ar, crude lauroyl chloride (25 mmol) and sodium hydroxide (1.0 g dissolved in 1.6 mL of water) were simultaneously added over 4 h while maintaining the pH at ~10. The reaction mixture was then allowed to stir overnight at room temperature. Water was added (10 mL, final pH ~ 9), and the solution was extracted with ethyl acetate (3 × 20 mL). The aqueous layer was treated with 1 M HCl to reach pH ~ 5, and extracted with ethyl acetate (3 × 20 mL). Finally, the aqueous layer was treated with 6 M NaOH to reach pH ~ 14, and extracted with ethyl acetate (3 × 20 mL). The aqueous layer was deposited in an Erlenmeyer flask and was allowed to crystallize overnight. The white solid was filtered and recrystallized from isopropanol to afford sodium *N*-lauroyl-*N*-methyltaurate (4.8 g, 56%). $^{1}$H-NMR (300 MHz, DMSO-$d_6$, 1:1 mixture of amide rotamers, Figure A.1a) δ 0.85 (t, $J$ = 6.7 Hz, 3H), 1.24 (m, 16H), 1.45 (m, 2H), 2.18-2.29 (m, 2H), 2.57-2.66 (m, 2H), 2.74 (s, 3H, rotamer A), 2.91 (s, 3H, rotamer B), 3.44-3.52 (m, 2H) ppm. $^{13}$C-NMR (75 MHz, MeOH-$d_4$, 1:1 mixture of amide rotamers, Figure A2) δ 14.5 ($CH_3$), 23.8 ($CH_2$), 26.2 ($CH_2$), 26.7 ($CH_2$), 30.5 ($CH_2$), 30.6 ($CH_2$), 30.7 ($CH_2$), 30.8 ($CH_2$), 33.1 ($CH_2$), 33.8 ($CH_3$), 33.9 ($CH_2$), 34.5, ($CH_2$), 36.7 ($CH_3$), 45.5 ($CH_2$), 47.2 ($CH_2$), 49.3 ($CH_2$), 49.9 ($CH_3$), 50.3 ($CH_2$), 175.9 (C), 176.2 (C) ppm. ESI (negative mode) m/z = 320.1890 (calc. for $C_{15}H_{30}NO_4S$, m/z = 320.1896).

**A.3. Synthesis of sodium N-d$_{23}$-lauroyl-N-methyltaurate.**

A similar procedure was followed, using d$_{23}$-dodecanoinc acid as starting material. $^{1}$H-NMR (300 MHz, DMSO, Figure A.1b) δ 2.57-2.66 (m, 2H), 2.84 (s, 3H), 3.44-3.52 (m, 2H) ppm. ESI (negative mode) m/z = 320.1940 (calc. for $C_{15}H_{30}NO_4S$, m/z = 320.1946). ESI (negative mode) m/z = 343.3344 (calc. for $C_{15}H_7D_{23}NO_4S$, m/z = 320.3339).

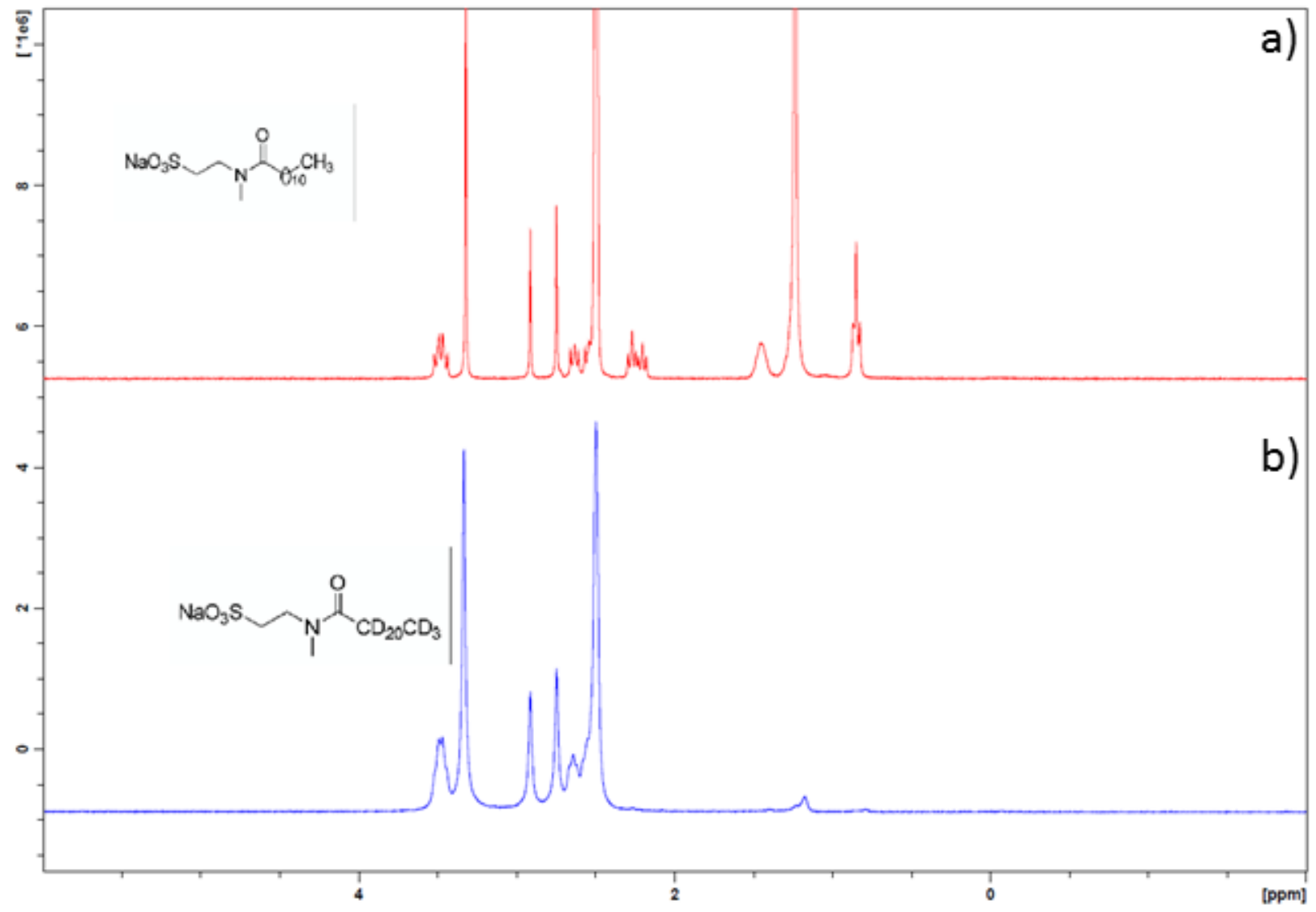

**Figure A.1.** $^{1}$H-NMR spectra (300 MHz) for hydrogenous (a) and deuterated (b) SLMT obtained in d$_6$-DMSO and DMSO, respectively.

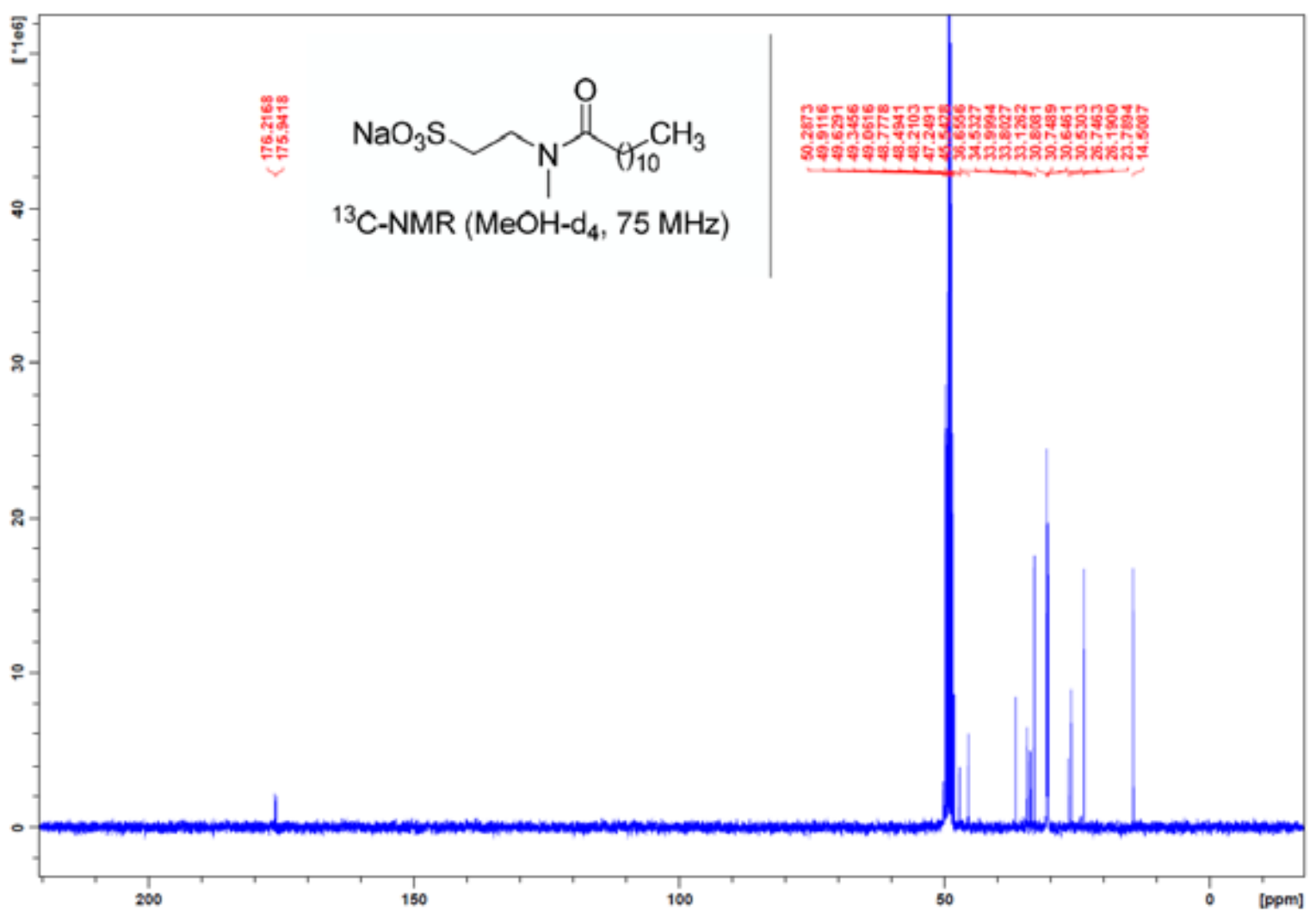

**Figure A.2.** $^{13}$C-NMR spectra (75 MHz) for hydrogenous SLMT obtained in $d_4$-MeOH.

## ReferencesUncategorized References